\documentclass{pas}
\usepackage{float}
\usepackage[varg]{txfonts}
\usepackage{graphicx}
\usepackage[utf8]{inputenc}

\renewcommand{\figcaptionnumfont}{\bfseries}
\renewcommand{\figcaptiondescfont}{\small}
\renewcommand{\figcaptionfont}{\small} 
\renewcommand{\tablecaptionnumfont}{\bfseries}
\renewcommand{\tablecaptionfont}{\small} 
\renewcommand{\tablefont}{\small}
\renewcommand\footnotesize{\small}

\begin{document}

\lefttitle{Cloud-scale simulations with cosmological ICs}
\righttitle{R. Ramesh, D. Nelson, D. Fielding \& M. Br\"{u}ggen}

\jnlPage{}{}
\jnlDoiYr{}
\doival{}

\articletitt{Research Paper}

\title{Zooming in on the circumgalactic medium with GIBLE: \\Cloud-scale simulations with cosmological initial conditions
}

\author{Rahul Ramesh$^{1,2}$, Dylan Nelson$^{2,3}$, Drummond Fielding$^{4}$ and Marcus Br\"{u}ggen$^{5}$}

\affil{$^1$Kavli IPMU (WPI), UTIAS, The University of Tokyo, Kashiwa, Chiba 277-8583, Japan\\
$^2$Universität Heidelberg, Zentrum für Astronomie, ITA, Albert-Ueberle-Str. 2, 69120 Heidelberg, Germany\\
$^3$Universität Heidelberg, Interdisziplinäres Zentrum für Wissenschaftliches Rechnen, INF 205, 69120 Heidelberg, Germany\\
$^4$Department of Physics, New York University, 726 Broadway, New York, NY 10003, USA\\
$^5$Hamburg University, Hamburger Sternwarte, Gojenbergsweg 112, 21029 Hamburg, Germany}
\corresp{Rahul Ramesh, Email: rramesh@g.ecc.u-tokyo.ac.jp}

\citeauth{}

\history{}

\begin{abstract}
We conduct simulations of $\sim$\,kpc-scale cool clouds in the circumgalactic medium (CGM), using initial conditions sampled from a highly resolved cosmological magneto-hydrodynamical zoom-in of a Milky Way–like galaxy. We select ten distinct cold clouds with masses of $m_{\rm{cloud}}$\,$\sim$\,$10^{4.5-5}$\,M$_\odot$, originally resolved at a mass resolution of $m_{\rm{gas}}$\,$\sim$\,$200$\,M$_\odot$. To further resolve small-scale features and physics, we implement a targeted refinement scheme within spherical regions co-moving with each cloud, thereby boosting the local mass resolution by a factor of 1000, reaching $m_{\rm{gas}}$\,$\sim$\,$0.2$\,M$_\odot$ (spatial resolution, $r_{\rm{gas,cloud}}$\,$\sim$\,$O(\rm{pc})$). The selected clouds have diverse properties, across a broad parameter space, resulting in heterogeneous evolution. For the clouds we study, radiative cooling is the dominant physical process enabling cloud survival, while magnetic fields play a comparatively smaller role. The motion of these clouds is governed not only by drag forces that decelerate them, but also by acceleration from momentum exchange with the complex background velocity field, which can cause them to move faster than ballistic projectiles set by their initial velocities. Our results suggest that the non-trivial details of realistic cosmological initial conditions -- specifically the complex density, temperature, and velocity fields -- may play an important role in subsequent cloud evolution, and that sampling the output of an existing large-scale simulation provides a self-consistent approach to capture these effects without ad hoc assumptions.
\end{abstract}


\maketitle

\section{Introduction}\label{sec:intro}
In a $\Lambda$CDM universe, galaxies form within halos of matter rather than in isolation. While such halos are dominated by dark matter in terms of mass, they also host gas. Depending on the mass of the system, these diffuse galactic atmospheres are commonly referred to as the circumgalactic medium (CGM; galaxy-scale halos), the intragroup medium (IGrM; group-scale halos), or the intracluster medium (ICM; cluster-scale halos). In all cases, theoretical models and observations suggest that these gaseous reservoirs are multi-phase, with cold clumps of gas (T\,$\sim$\,$10^4$\,K) embedded within a volume-filling warm-hot phase, which can reach temperatures several orders of magnitude higher (see \citealt{donahue2022} for a recent review).

The spatial and temporal co-evolution of gas in such distinct phases has long posed a challenge for theoretical models. Hydrodynamic instabilities, such as the Kelvin–Helmholtz instability (KHI), develop at the interface between clouds and the ambient medium due to velocity shear, driving mixing and disruption of the cold phase \citep[e.g.][]{klein1994}. In addition, cold clouds are susceptible to compression by the thermal pressure of the surrounding hot gas \citep[][]{mckee1975} or evaporation through thermal conduction, which can remove mass from the cloud over time \citep[][]{mckee1977}. The latter is particularly important in galaxy clusters -- where the typical hot phase temperatures are T\,$\gtrsim$\,$10^7$\,K -- since the thermal conductivity in a fully ionized plasma scales sharply as $\sim$\,T$^{5/2}$ \citep{spitzer1962,cowie1977}.

These challenges have motivated a number of simulations and analytic models that investigate the lifetimes, disruption mechanisms, and survival of cold clouds in hot halos, as well as how cloud properties and physical processes shape their evolution.

Starting from relatively small spatial scales, KHI setups -- where gases of two different densities are initialized with a velocity contrast at the interface that separates them \citep[e.g.][]{mcnally2012,lecoanet2016} -- provide a controlled framework to study how shear flows affect the boundary between clouds and their surroundings. In their simplest form, where only radiative cooling is coupled with hydrodynamics, these simulations are characterized by a small set of parameters, making them an ideal testbed for exploring phase interactions at small scales. However, they can also readily incorporate more complex physics, such as non-equilibrium ionization (NEI) effects \citep{kwak2010} or magnetic fields \citep{esquivel2006,ji2019}.

In recent years, these setups have played a pivotal role in quantifying the physics of radiative turbulent mixing layers -- the interface zones surrounding clouds that host intermediate-temperature gas \citep[][]{begelman1990}. For instance, \cite{fielding2020} show that cooling losses in these layers are balanced by the inflow of hot, high-enthalpy gas, and that the cooling occurs within a boundary layer whose surface is highly irregular and fractal-like. They further demonstrate that this complex geometry sets the scale-dependence of inflow velocity and cooling rate in the mixing layer. Note, however, that recent work by \cite{sharma2025} argues that a dip in thermal pressure within the layer may be sustained by a positive vertical compressive stress, suggesting an additional dynamically important contribution from turbulent stresses compared to a purely advective–radiative framework (see also \citealt{chen2023}).

At scales beyond the immediate cloud-background interface, cloud crushing simulations make it possible to investigate their evolution and survival as they move through their ambient medium. Early studies of this kind typically modeled initially spherical\footnote{Note, however, that other simple cloud geometries such as ellipsoids have also been explored \citep[e.g.][]{xu1995}.} clouds with uniform temperature and density embedded in a homogeneous, hot background flow, assuming a constant velocity shear between the two components \citep[e.g.][]{stone1992,nakamura2006,marinacci2010,armillotta2017}.

Although these setups are relatively simple, they have significantly advanced our understanding of cloud growth \citep[e.g.][]{bruggen2016}. Using hydrodynamic simulations that include radiative cooling, \cite{gronke2018} show that when the cooling time of the interface gas is shorter than the timescale over which clouds are shredded by hydrodynamic instabilities, they experience a net influx of cold mass from the wind. As a result, they are likely to survive, which can also be translated into a minimum initial size criterion for their long-term persistence \citep[see also][]{gronke2020}.

The picture turns increasingly complex as additional physical processes are considered. \cite{mccourt2015} show that the presence of magnetic fields -- particularly when field lines at the interface are draped around the cloud \citep{dursi2008} -- can enhance the drag force, causing the cloud to begin co-moving with the wind more quickly and effectively reducing the time it is exposed to shredding \citep[see also][]{sparre2020}. \cite{bruggen2023} further demonstrate that the topology of magnetic field lines may be important in the presence of anisotropic thermal conduction and can contribute to the damping of instabilities.

There have also been numerous efforts to move beyond overly simplified initial conditions in these simulations. These include, but are not limited to, beginning with clouds that have more complex geometries \citep{cooper2009,yao2025}, adopting non-homogeneous density profiles within the cloud and/or the surrounding medium \citep{bandabarragan2020,jung2023}, injecting turbulence into the ambient velocity field \citep{ghosh2025}, and varying the thermodynamic properties of the background as the cloud progresses, in order to mimic the changing conditions encountered as clouds traverse their host galaxy or halo \citep{dutta2025,hidalgopineda2025}.

Other common idealized setups for studies of this kind include tall-box simulations, which model a patch of the interstellar medium (ISM) along with an extended region perpendicular to the disk \citep[e.g.][]{girichidis2016,kim2017}, and isolated galaxy simulations, which typically cover a larger volume \citep[e.g.][]{fielding2017,smith2021,fournier2024}. In both approaches, clouds do not need to be initialized individually; instead, they are naturally `seeded' as feedback processes inject energy into the surrounding ISM and drive gas outwards.

For example, \cite{tan2024} use a tall-box simulation to study the formation, evolution, and physical properties of cool clouds that naturally emerge within hot galactic outflows. They show that these clouds in the inner CGM originate through the dynamical fragmentation of the ISM and exhibit complex morphologies, turbulent support, and mass growth rates consistent with models of turbulent radiative mixing layers. Similarly, \cite{warren2025} use the CGOLS simulations \citep{schneider2018} to study large populations of cool clouds in galactic outflows, finding that their mass and velocity distributions are consistent with turbulent fragmentation and motions, with cloud survival largely predictable from recently proposed theoretical criteria.

Despite their strengths, the various numerical experiments discussed above share some intrinsic limitations. Firstly, their results often depend strongly on the chosen initial conditions, which are typically generated in an ad hoc manner, and the explored parameter space may not capture the full range of realistic cloud properties, limiting generalization \citep{jennings2022}. Moreover, even with sophisticated initial setups, reproducing the full structural and dynamical diversity of a realistic CGM remains challenging. In the case of tall-box and isolated galaxy simulations, these limitations are partially mitigated, but such models primarily capture the physics of the disk–halo interface or inner halo, without extending to the outer halo or accounting for large-scale environmental influences such as gas accretion or interactions with satellite galaxies.

Cosmological hydrodynamical simulations provide a complementary avenue for exploring the CGM in a more self-consistent context. By following the assembly of galaxies and their gaseous halos across cosmic time, they naturally capture the complex interplay between galactic outflows \citep[e.g.][]{oppenheimer2008,ayromlou2023}, large-scale gas accretion \citep[e.g.][]{keres2005,nelson2013}, and satellite interactions \citep[e.g.][]{olano2008,rohr2023}. Although their resolution remains lower than that of idealized counterparts, ongoing developments are steadily narrowing this gap, enabling the study of small-scale gas structures within a fully evolving $\Lambda$CDM framework \citep[see Figure~1 of][]{pillepich2023}.

Cosmological zoom-in simulations, in particular those that employ targeted refinement in specific regions -- like the CGM -- make it possible to boost numerical resolution in a computationally efficient manner \citep[e.g.][]{suresh2019,voort2019,rey2025}. Such setups have recently been used to quantify the properties of cool clouds \citep[][]{hummels2019,augustin2025}, investigate their formation channels \citep[][]{lucchini2024}, and follow their long-term evolution over timescales of $\gtrsim$\,Gyr \citep[][]{ramesh2025}, a duration accessible owing to the fully cosmological nature of these runs.

Although these projects have begun to `bridge' the resolution gap between standard cosmological simulations and their idealized counterparts, they still fall short of resolving clouds at comparable levels of detail -- an important limitation, as resolution is known to influence cloud evolution \citep[e.g.][]{yirak2010,pittard2016}. The question therefore remains: can this bridge be extended further?

In this work, we take a step forward and sample, for the first time, initial conditions for a cloud-scale simulation directly from a high resolution cosmological run, and subsequently perform simulations analogous to existing idealized setups. As described in detail in Section~\ref{sec:methods}, we implement an additional refinement scheme to ultra-refine gas in and around the cloud, boosting the resolution to levels approaching those of modern cloud-crushing simulations. This technique provides a compromise between fully cosmological simulations that capture large-scale evolution at low resolution and idealized cloud simulations that achieve higher resolution but lack cosmological context. Conceptually, this strategy is similar to ongoing efforts that zoom in on the accretion disks of supermassive black holes using initial conditions drawn from cosmological simulations \citep[][]{hopkins2024}.

The rest of the paper is organized as follows: in Section~\ref{sec:methods}, we describe the procedure for sampling initial conditions, the targeted refinement scheme used to enhance resolution around clouds, and the definitions and physical processes adopted throughout. Results are presented and discussed in Section~\ref{sec:results}, and conclusions are summarized in Section~\ref{sec:summary}.

\section{Methods}\label{sec:methods}

\subsection{Extracting initial conditions}\label{ssec:extract_ics}
For our initial conditions, we sample the $z$\,$=$\,$0$ snapshot volume from the highest-resolution run of Project GIBLE \citep[][]{ramesh2024b}, a cosmological magnetohydrodynamical zoom-in simulation of a Milky Way–like galaxy that employs super-Lagrangian refinement to achieve a mass resolution of $\sim$\,$225$\,M$_\odot$ in the circumgalactic medium\footnote{This is defined as the region bounded between 0.15\,R$_{\rm{200c}}$ and R$_{\rm{200c}}$ (virial radius).}. 

The details of the simulation setup are described extensively in \citet{ramesh2024a}; we here provide a brief summary. Run with the \textsc{arepo} moving-mesh code \citep{springel2010}, Project GIBLE adopts the IllustrisTNG galaxy formation framework \citep{weinberger2017,pillepich2018} to model the key baryonic processes governing galaxy assembly and evolution. These include primordial and metal-line cooling in the presence of a time-varying ultraviolet background (UVB), star formation, stellar evolution and chemical enrichment, the formation of supermassive black holes (SMBHs), feedback from both supernovae and SMBH activity, and ideal magnetohydrodynamics \citep{pakmor2011,pakmor2014}.

The first step in constructing the initial conditions is to identify cold gas clouds within the CGM. Following \cite{nelson2020,ramesh2023b}, we define these as contiguous groups of Voronoi cells with temperatures T\,$\leq$\,$10^{4.5}$\,K. We further restrict our selection to gas cells that are not gravitationally bound to any satellite galaxies, as determined by the substructure identification algorithm \textsc{subfind} \citep{springel2001}.

Next, we apply a set of criteria to select a sub-sample of clouds from the identified population: 

\begin{enumerate}

\item \textbf{Mass threshold:} To ensure that each cloud is sufficiently resolved, we retain only those with masses $\rm{M_{cld}}$\,$>$\,$10^{4.5}$\,M$_\odot$, corresponding to $\gtrsim$\,$150$ gas cells per cloud.

\item \textbf{Distance cutoff:} We require the galactocentric distance of each cloud to lie within $0.3$\,R$_{\rm{200c}}$ and $0.85$\,R$_{\rm{200c}}$. This ensures that the extracted regions remain well within the high-resolution domain of the parent cosmological simulation, thereby avoiding contamination from gas cells at significantly lower resolution.

\item \textbf{Isolation criterion:} To simplify the analysis and facilitate physical interpretation, we select clouds that are relatively isolated. To do so, we first compute the cloud-crushing timescale \citep[][]{klein1994}
\begin{equation}
    \rm{t_{cc}} = \chi^{1/2} \, \rm{R_{cld}} / \rm{v_{rel}},
\end{equation}
where $\chi$ and $\rm{v_{rel}}$ are the density and velocity contrasts between the cloud and its surrounding background (see Section~\ref{ssec:def} for detailed definitions), respectively. $\rm{R_{cld}}$ is the cloud radius, defined as the geometric mean of the three semi-axes of the ellipsoid that best fits all vertices of the outer layer of the cloud. We then estimate the distance a cloud would travel during this time if it were a ballistic projectile with its initial velocity ${\overrightarrow{\rm{v}_{\rm{cld}}}}$:
\begin{equation}
    \rm{d_{cld, t_{cc}}} = |{\overrightarrow{\rm{v}_{\rm{cld}}}}| \, \rm{t_{cc}}.
\end{equation}
Next, we compute the maximum mass of neighboring clouds within a distance of $3$\,$\rm{d_{cld, t_{cc}}}$ from the cloud center ($\rm{M_{max,neigh}}$) and discard all clouds with $\frac{\rm{M_{max,neigh}}}{\rm{M_{cld}}}$\,$\geq$\,$0.25$. Although this selection facilitates a cleaner interpretation of individual cloud evolution, it does so at the expense of capturing direct cloud–cloud interactions \citep[e.g.][]{williams2022}. Nevertheless, as discussed in later sections, two of the selected systems do experience mergers, indicating that this isolation criterion is not perfectly restrictive, particularly because it does not take the direction of motion of the neighboring clouds into account.

\end{enumerate}

Applying the above criteria yields a total of 13 distinct clouds. We then rank these in ascending order of their $\rm{t_{cc}}$ and select the first ten. This choice both minimizes computational cost and ensures that the extracted setups remain physically meaningful within a cosmological context -- i.e. running these simulations for $\gg$\,$O(100 \rm{Myr})$ would no longer be realistic once galaxy- and halo-scale physics are not included.

\begin{figure*}
    \centering
    \includegraphics[width=18cm]{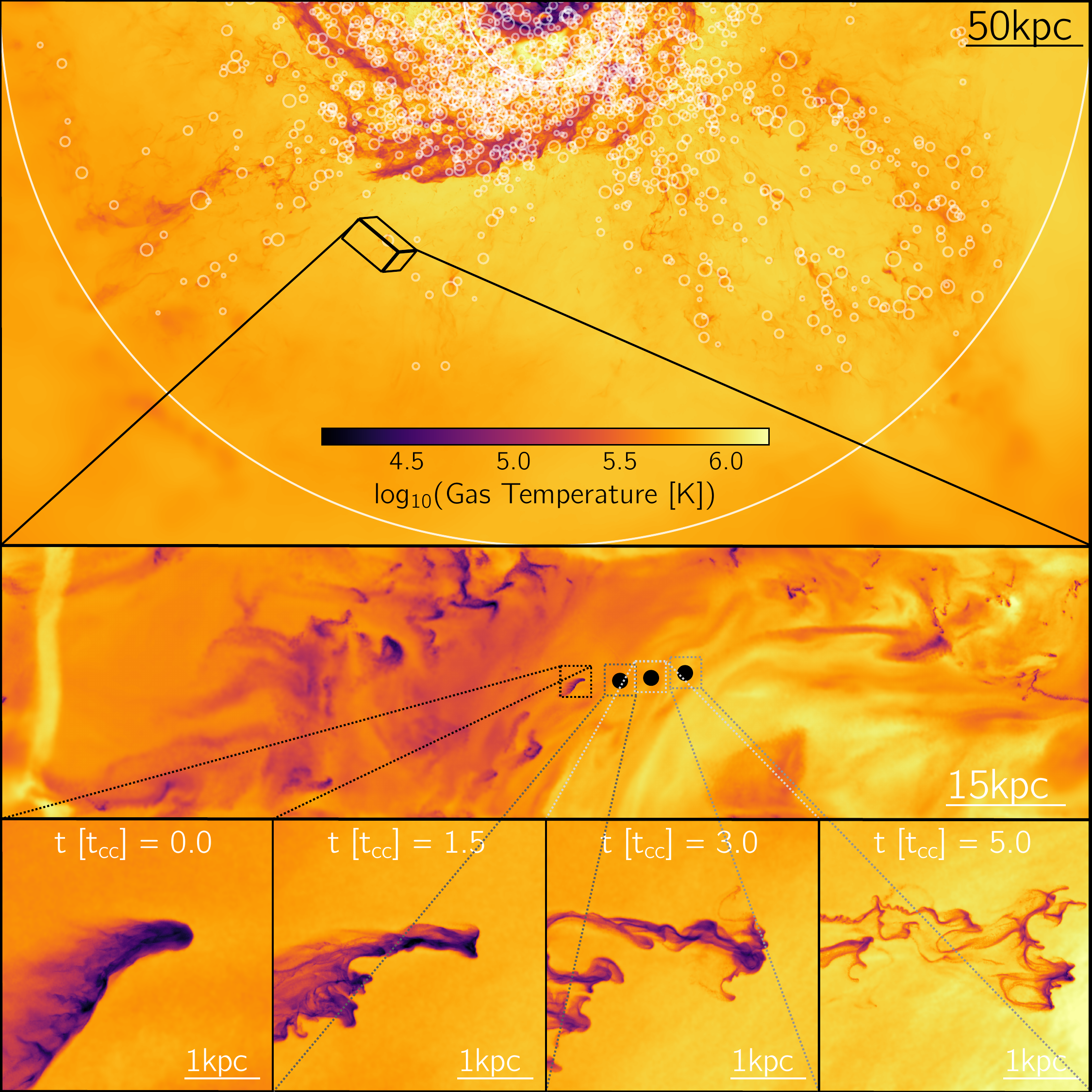}
    \caption{An exemplary visualization of the technique explored in this work. The top panel shows a halo-scale projection of gas temperature, with the image stretching $\pm$\,R$_{\rm{200c}}$ along the $x$ and $z$ axes, and R$_{\rm{200c}}$ along the $y$ axis. The white semi-circle is drawn at R$_{\rm{200c}}$, and the smaller circles at the positions of all clouds above a mass threshold of M$_{\rm{cl}}$\,$\gtrsim$\,$10^5$\,M$_\odot$, with the radii of the circles scaled with the size of clouds. The cuboid represents a sample region (not to scale), centered on a cloud of radius R$_{\rm{cl}}$, extracted to be used as an initial condition for the cloud-scale simulation. The center panel shows a thin $\pm$\,R$_{\rm{cl}}$ projection of this volume. The chosen cloud (\#4; see Table~\ref{table:cloud_props}) is highlighted by the dotted black box, and a zoomed-in projection is shown in the lower-left panel. The black dots in the center panel are drawn at the positions of the main descendants at future times (with a small horizontal offset to ensure they do not overlap), and the corresponding insets in the lower panel show analogous temperature-projections. While most of the volume in the extracted region is simulated at a mass resolution of $\sim$\,$200$\,M$_\odot$ (center panel), a targeted spherical refinement region co-moving with the cloud boosts the local resolution to $\sim$\,$0.2$\,M$_\odot$ (as seen in the lower panels). The elongated side of the cuboid is aligned with the cloud’s initial direction of motion, and all lower panels are rotated so that the cloud’s instantaneous mean velocity points to the right. This example shows how the method enables setting cloud-scale initial conditions while retaining the complexity of the surrounding CGM.}
    \label{fig:mainIntroVis}
\end{figure*}

\begin{table*}
\centering
\begin{tabular}{||c|c|c|c|c|c|c|c|c|c|c|c|c||} 
 \hline
 Cloud & M$_{\rm{cld}}$ & R$_{\rm{cld}}$ & v$_{\rm{rel}}$ & $\chi$ & v$_{\rm{rad}}$ & Gal. Dist. & $\epsilon$ & A$_{\rm{cld}}$ & $\mu_{\rho, \rm{bck}}$ & $\mu_{\rm{T, bck}}$ & $\sigma_{\rho, \rm{bck}}$ & $\sigma_{\rm{T, bck}}$ \\ 
 \# & [M$_\odot$] & [kpc] & [km/s] & & [km/s] & [R$_{\rm{200c}}$] & & [kpc$^2$] & [M$_\odot$/kpc$^{3}$] & [K] & [M$_\odot$/kpc$^{3}$] & [K] \\ [.5ex] 
 \hline\hline 
 1  & $10^{5.28}$ & 0.68 & 51.18 & 37.31 & +2.11 & 0.58 & 4.21 & 15.17 & $10^{3.41}$ & $10^{5.92}$ & $10^{3.45}$ & $10^{5.81}$ \\ 
 2  & $10^{4.99}$ & 0.74 & 67.01 & 29.33 & +3.59 & 0.56 & 4.76 & 31.19 & $10^{3.61}$ & $10^{5.84}$ & $10^{3.69}$ & $10^{5.87}$ \\
 3  & $10^{4.68}$ & 0.46 & 47.61 & 18.49 & +4.09 & 0.67 & 2.75 & 14.96 & $10^{4.08}$ & $10^{5.68}$ & $10^{4.19}$ & $10^{5.77}$ \\
 4  & $10^{4.59}$ & 0.39 & 59.01 & 42.88 & -45.31 & 0.45 & 4.07 & 14.09 & $10^{3.71}$ & $10^{5.93}$ & $10^{3.79}$ & $10^{5.81}$ \\
 5  & $10^{4.75}$ & 0.72 & 68.78 & 29.19 & -48.28 & 0.65 & 2.33 & 22.21 & $10^{3.77}$ & $10^{5.78}$ & $10^{3.74}$ & $10^{5.77}$ \\
 6  & $10^{4.52}$ & 0.45 & 43.19 & 20.25 & -3.11 & 0.54 & 2.16 & 13.09 & $10^{3.69}$ & $10^{5.71}$ & $10^{3.59}$ & $10^{5.59}$ \\
 7  & $10^{4.50}$ & 0.51 & 37.98 & 21.97 & +14.68 & 0.49 & 6.19 & 29.04 & $10^{3.71}$ & $10^{5.65}$ & $10^{3.72}$ & $10^{5.77}$ \\
 8  & $10^{4.51}$ & 0.22 & 45.91 & 57.19 & -2.99 & 0.32 & 2.54 & 3.70 & $10^{3.77}$ & $10^{5.89}$ & $10^{3.47}$ & $10^{5.67}$ \\
 9  & $10^{4.68}$ & 0.79 & 55.49 & 16.37 & +46.76 & 0.79 & 3.42 & 11.95 & $10^{3.56}$ & $10^{5.68}$ & $10^{3.48}$ & $10^{5.73}$ \\
 10 & $10^{4.83}$ & 0.81 & 65.11 & 12.77 & -62.73 & 0.73 & 2.84 & 24.86 & $10^{3.31}$ & $10^{5.61}$ & $10^{3.25}$ & $10^{5.68}$ \\ [.5ex] 
 \hline
\end{tabular}
\vspace{0.2cm}
\caption{A summary of select few properties of the sampled clouds and their ambient surroundings at t\,$=$\,$0$. From left to right, the columns list: cloud number (a common index with other figures); cloud mass (M$_{\rm{cld}}$); cloud radius (R$_{\rm{cld}}$); relative velocity (v$_{\rm{rel}}$); overdensity ($\chi$); radial velocity (v$_{\rm{rad}}$); galactocentric distance; major-to-minor axis ratio ($\epsilon$); surface area (A$_{\rm{cld}}$); mean density ($\mu_{\rho, \rm{bck}}$) and temperature ($\mu_{\rm{T, bck}}$) of the ambient background; and the corresponding standard deviations ($\sigma_{\rho, \rm{bck}}$, $\sigma_{\rm{T, bck}}$). 
See main text for further details (Section~\ref{sec:results}).
}
\label{table:cloud_props}
\end{table*}

For each cloud, the size of the extracted simulation domain is defined by first computing a characteristic length scale,

\begin{equation}
    \rm{d_{box}} = \rm{max}(20\,\rm{R_{cld,max}}, 4\,\rm{d_{cld, t_{cc}}}),
\end{equation}

where $\rm{R_{cld,max}}$ is the length of the semi-major axis of the cloud. The final volume is then specified as [$\rm{L_x}$, $\rm{L_y}$, $\rm{L_z}$]\,$=$\,[$\pm$\,$4\,\rm{d_{box}}$, $\pm$\,$\rm{d_{box}}$, $\pm$\,$\rm{d_{box}}$], centered on the cloud and oriented such that the x-axis aligns with its instantaneous velocity vector. Overall, this provides a conservative measure of the region required to contain the cloud and its immediate surroundings during the simulation, which we always stop at $5$\,$\rm{t_{cc}}$.

Periodic boundary conditions are applied in all directions; the chosen domain size ensures that the cloud and its fragments do not interact with their own periodic images or repeatedly encounter the same background medium. This configuration also ensures that any numerical artifacts introduced by the assumed periodicity remain dynamically negligible.

The upper two panels of Figure~\ref{fig:mainIntroVis} illustrate the procedure described above. In the top panel, we show a halo-scale projection of the gas temperature of the parent GIBLE halo, with the image extending $\pm$\,R$_{\rm{200c}}$ along the $x$ and $z$ axes and R$_{\rm{200c}}$ along the $y$ axis. The semi-circle is drawn at R$_{\rm{200c}}$, while smaller circles mark the positions of all clouds above a mass threshold of M$_{\rm{cl}}$\,$\gtrsim$\,$10^5$\,M$_\odot$, with the circle radii scaled to the corresponding cloud sizes. The cuboid represents a sample region (not to scale), centered on a cloud of radius R$_{\rm{cl}}$, extracted for use as the initial condition in the cloud-scale simulation. The center panel shows a thin $\pm$\,R$_{\rm{cl}}$ projection of this volume. Note the bright vertical streak on the left edge of the panel: this is a numerical artifact introduced by the assumed periodicity, but the chosen extraction volume ensures that it does not affect our results, as discussed above, and this has been further verified by testing with larger boxes.

\subsection{Cloud-centric refinement scheme}\label{ssec:ref_scheme}
While the above extraction procedure provides initial conditions that are cosmologically motivated, the resolution on $\sim$\,kpc scales remains limited and does not compete with that of modern cloud-crushing simulations. We therefore implement a targeted cloud-centric refinement scheme to increase the local mass resolution, building upon the existing (de-)refinement scheme in \textsc{arepo} \citep[][]{springel2010}.

To achieve this, we employ a cell-flagging procedure to define the required refinement region, analogous to passive scalar based cloud-tracking techniques \citep[e.g.,][]{scannapieco2015}. Briefly, we flag every gas cell that initially belongs to the cloud. When a cell refines into two, both descendants inherit the flag. Conversely, when a cell de-refines and deposits its contents into neighboring cells, the flag is passed on to all of them. The center of the refinement region is then defined as the center-of-mass of all gas cells containing a non-zero flag. 

At each time step, all cells within a radius of $15$\,$\rm{R_{cld,max}}$\footnote{Note that this is the length of the semi-major axis computed at the initial conditions, and the refinement radius is kept fixed throughout the simulation. While we find this to be sufficient when simulations are evolved only to $5$\,$\rm{t_{cc}}$, a larger refinement region may be necessary for longer time-evolutions. Alternatively, it may be desirable to have an adaptive radius for the refinement region, although we do not experiment with this in the current work.} are refined to a target gas mass resolution, which we set to $\sim$\,$0.2$\,M$_\odot$ for the main body of this work (see Appendix~\ref{sec:app_res_conv} for convergence tests against lower-resolution runs). Note that we do not use an intermediate buffer region to smoothly vary the resolution with increasing distance from the refinement boundary, since this would make reaching the target resolution prohibitively expensive, even within the relatively small extracted volume. We have verified for a single test case that allowing a smooth transition in resolution outside the refinement region does not change the evolution of the cloud.

With the volume extracted (as described in Section~\ref{ssec:extract_ics}) and the cloud-based refinement established, we evolve the initial conditions with our fiducial physics setup (see Section~\ref{ssec:phys_proc}) for $\sim$\,$20$\,Myr for all clouds. This allows the cells to reach the target resolution and the mesh to regularize. We then adopt these configurations as the initial conditions for the subsequent evolution. The first panel of the bottom row of Figure~\ref{fig:mainIntroVis} shows a thin $\pm$\,R$_{\rm{cl}}$ zoomed-in projection centered on the cloud from the middle row, illustrating the temperature structure at an enhanced resolution of $m_{\rm{gas}}$\,$\sim$\,$0.2$\,M$_\odot$\footnote{The corresponding spatial resolutions are as high as $r_{\rm{gas,cloud}}$\,$\sim$\,$O(\rm{pc})$ within clouds, and a factor of a few coarser in the ambient gas (see also Figure~\ref{fig:app_inter_iden}).} at t\,$=$\,$0$. At this level, clouds with masses $\sim$\,$10^{4.5-5}$\,M$_\odot$ are resolved by $\sim$\,$50-60$ cells per cloud radius, comparable to high-resolution idealized cloud-crushing simulations.

\begin{figure*}
    \centering
    \includegraphics[width=18cm]{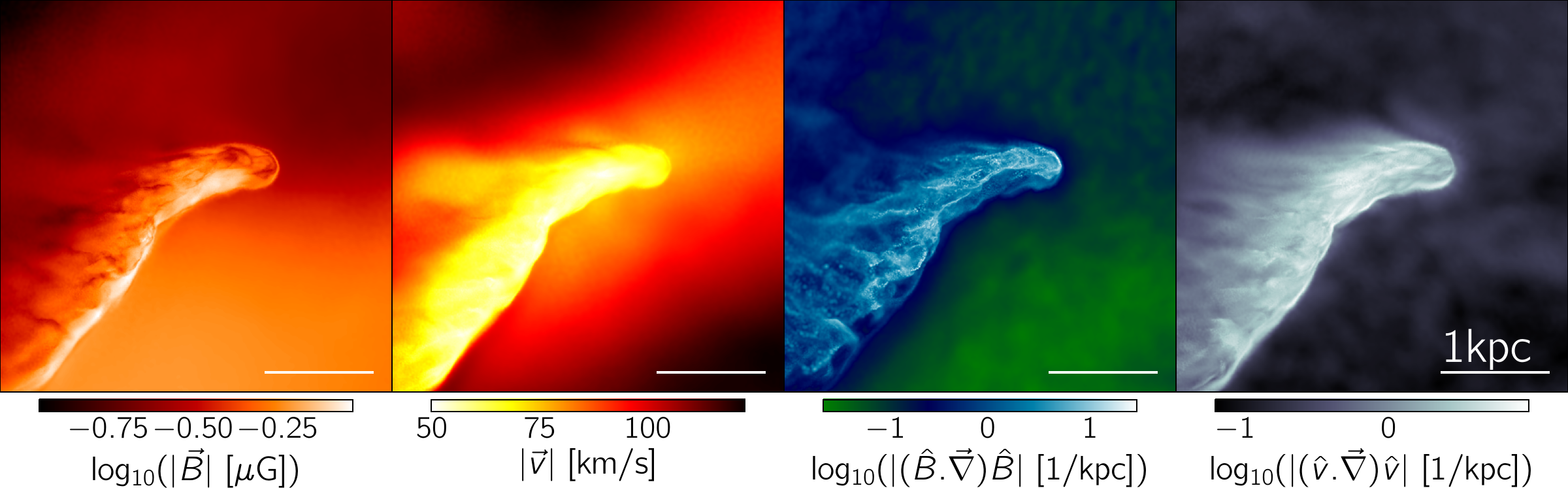}
    \caption{A visualization of the small-scale velocity and magnetic fields around the same cloud shown in Figure~\ref{fig:mainIntroVis}. From left to right, panels show thin $\pm$\,R$_{\rm{cl}}$ projections of the magnitudes of the magnetic field, the velocity, and their respective curvatures. All panels are oriented so that the cloud’s mean velocity points to the right. Both the magnetic and velocity fields exhibit substantial structure on these scales: their magnitudes vary throughout the cloud and background, and the field directions are complex and non-uniform.}
    \label{fig:propVis}
\end{figure*}

\subsection{Physical processes}\label{ssec:phys_proc}
To remain as consistent as possible with the GIBLE simulations from which we extract initial conditions, we adopt the same physics modules for our fiducial setup, which include:
(a) primordial \citep{katz1996} and metal-line \citep{wiersma2009} cooling in the presence of a $z$\,$=$\,$0$ UVB \citep{fg2009}, with corrections for self-shielding at high physical densities \citep{rahmati2013}; and
(b) the evolution of magnetic fields following the equations of ideal MHD \citep{pakmor2011}, with the \citealt{powell1999} scheme used for divergence cleaning.

In addition, for a subset of clouds, we run three physics variations, starting from the generated initial conditions at t\,$=$\,$0$: (i) `Pure Hydro', in which both (a) and (b) are disabled; (ii) `No Cooling', in which (a) is disabled; and (iii) `No B', in which (b) is disabled. When magnetic fields are turned off, the corresponding magnetic energy is added to the thermal energy reservoir, and the total pressure remains roughly unchanged as a result \citep[see also][]{nelson2020}.

\subsection{Cloud-related definitions}\label{ssec:def}
Here we summarize definitions related to clouds that are used throughout this work:
\begin{enumerate}
    \item \textbf{Cloud Interface and Backgrounds:} 
    Motivated by past studies that have discussed the presence of an intermediate warm-phase layer sandwiched between clouds and their hot backgrounds \citep{gronke2018,fielding2020,nelson2020}, we define the interface as the \textit{transition} region extending outward from the cloud until the temperature is $95\%$ of its asymptotic value; we consider the latter as the background. As discussed in greater detail in Appendix~\ref{sec:app_inter_iden}, this is identified through the geometric connectivity of the Voronoi tessellation via an algorithm that adapts to the arbitrary shape of any given cloud.
    \item \textbf{Cloud Descendants:} As we evolve our extracted volumes forward, we save snapshots at intervals of $\sim$\,$0.1$\,$\rm{t_{cc}}$, resulting in a total of 50 snapshots. At each snapshot, we run our cloud-finding algorithm (see Section~\ref{ssec:extract_ics}). Every cloud containing at least one gas cell with a non-zero refinement flag is tagged as a \emph{descendant}, with the most massive designated as the \emph{main descendant}. This procedure is largely identical to the cloud-linking algorithm employed in \cite{ramesh2025}, but without the use of Monte Carlo tracers \citep{genel2013,nelson2013}.
\end{enumerate}

\section{Results}\label{sec:results}
We begin by examining in more detail the initial state of the same cloud shown in Figure~\ref{fig:mainIntroVis}, focusing on the velocity and magnetic-field line structures to highlight the non-trivial features captured by our method. From left to right, Figure~\ref{fig:propVis} shows thin $\pm$\,R$_{\rm{cl}}$ projections of the magnitudes of the magnetic field, the velocity, and their respective curvatures. The curvature values as defined here represent the inverse of the local radius of curvature of the field lines; smaller values therefore correspond to field lines that are locally straighter \citep{shen2003,boozer2005}. All panels are rotated such that the instantaneous velocity vector of the cloud points to the right.

\begin{figure*}
    \centering
    \includegraphics[width=18cm]{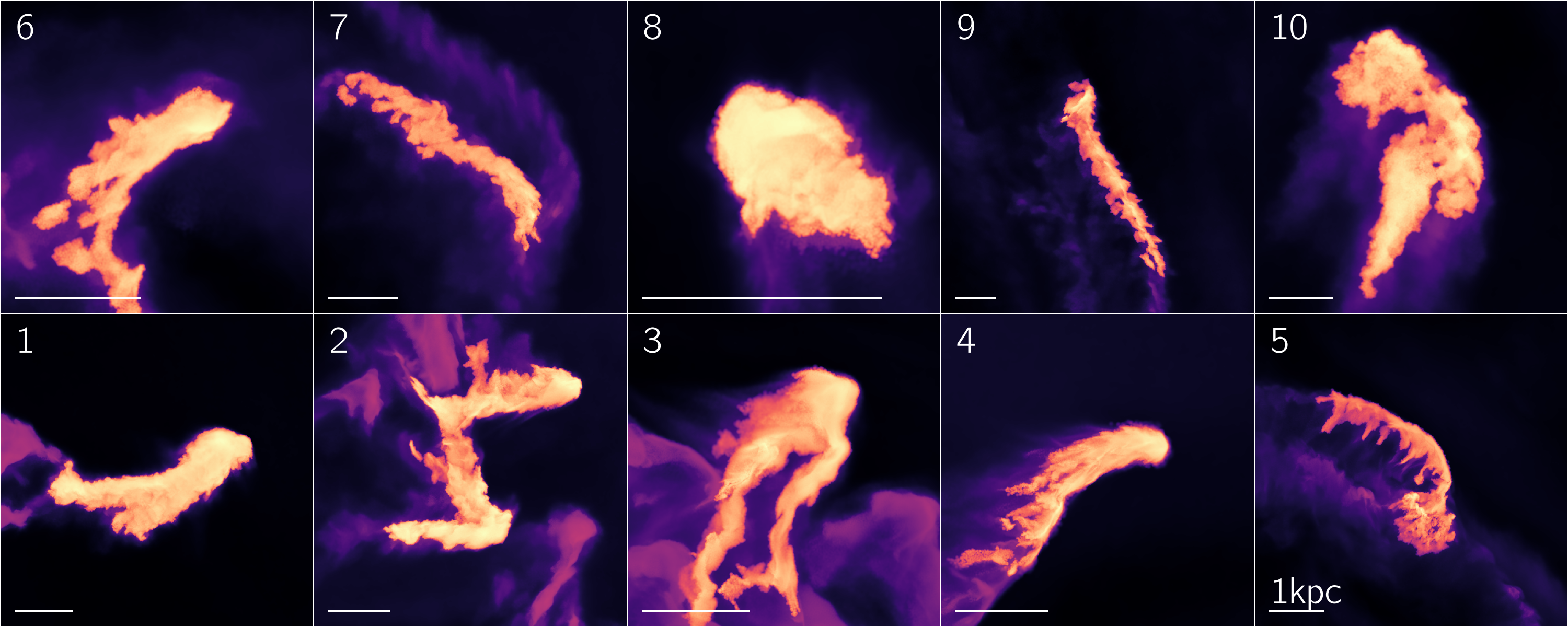}
    \includegraphics[width=7cm]{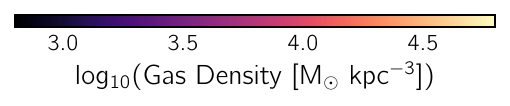}
    \caption{A gallery visualizing the initial state of the ten clouds selected for our sample, with all oriented so that their mean velocity vector points to the right. Colors indicate the mean gas density in a thin projection, and scale bars in all panels correspond to $1$\,kpc. Each cloud is unique in its initial morphology and exhibits distinct internal density distributions. Key properties of these clouds are summarized in Table~\ref{table:cloud_props}. This figure highlights the diversity of initial conditions captured by our selection, which underlies the variety of subsequent cloud evolution.}
    \label{fig:ic_gallery}
\end{figure*}

The magnetic field is not uniform throughout the cloud volume, but is instead more strongly magnetized along the leading, upstream edge \citep{dursi2008}. Additionally, within the body of the cloud, there are distinct `streaks' that are slightly less magnetized than the surrounding gas. While such small-scale features are largely absent in the background gas, a clear gradient in field strength is visible across the cloud, oriented perpendicular to its velocity vector. The magnetic-field curvature is generally larger inside the cloud than in the background; interestingly, the regions of weaker magnetization broadly coincide with higher curvature values \citep[see also][]{schekochihin2004}. 

Although small variations in both the magnitude and curvature of the velocity field are present within the cloud, it is overall moving as a coherent entity at a well-defined velocity. Downstream of the cloud, a patch of gas exhibits an intermediate velocity between that of the cloud and the background, indicating ongoing mixing or entrainment \citep{gronke2018}. The background flow is clearly complex: the cloud is embedded in a diagonal wake of gas extending from the bottom-left to the top-right, which itself is moving relative to the surrounding background at a different velocity. Overall, this level of structure is challenging to initialize with conventional, ad hoc procedures, highlighting a key advantage of our approach.

Transitioning from a single cloud to the entire sample, Figure~\ref{fig:ic_gallery} visualizes a gallery of the ten different clouds we consider in this work. Each panel displays a thin $\pm$\,R$_{\rm{cl}}$ projection of gas density at t\,$=$\,$0$, rotated such that the instantaneous velocity vector of the cloud points to the right. All scale bars correspond to a length of 1\,kpc. 

The clouds exhibit clear diversity in their initial morphologies and internal density structures. While most appear to form a single coherent object, some seem to represent the superposition of two smaller clouds that have recently undergone a merger (clouds 2 and 10; see also \citealt{ramesh2025}). The density field of the surrounding medium also varies from cloud to cloud: some show strong density inhomogeneities, whereas others appear relatively smooth and relaxed.

In Table~\ref{table:cloud_props}, we summarize a number of properties corresponding to the initial state of these clouds. From left to right, the columns list:
\begin{enumerate}
\item Cloud number, a common index with other figures.
\item M$_{\rm{cld}}$, the sum of the masses of all gas cells that comprise the cloud.
\item R$_{\rm{cld}}$, the cloud radius, defined as the geometric mean of the three semi-axes of the ellipsoid that best fits all vertices of the cloud’s outer layer.
\item v$_{\rm{rel}}$, the relative velocity between the cloud and the background.
\item $\chi$, the overdensity of the cloud with respect to the background.
\item v$_{\rm{rad}}$, the radial velocity of the cloud in the parent halo.
\item the galactocentric distance of the cloud in the parent halo.
\item $\epsilon$, the major-to-minor axis ratio of the ellipsoid that best fits all vertices of the cloud’s outer layer.
\item A$_{\rm{cld}}$, the surface area of the cloud.
\item $\mu_{\rho, \rm{bck}}$, the mean density of all gas cells that comprise the background.
\item $\mu_{\rm{T, bck}}$, the mean temperature of all gas cells that comprise the background.
\item $\sigma_{\rho, \rm{bck}}$, the standard deviation of the density of all gas cells that comprise the background.
\item $\sigma_{\rm{T, bck}}$, the standard deviation of the temperature of all gas cells that comprise the background.
\end{enumerate}

Note that the radial velocity and galactocentric distance of the clouds are computed at the time of extraction from the parent cosmological simulation. All other quantities are defined at t\,$=$\,$0$ of the subsequent runs (see Section~\ref{ssec:ref_scheme}).

Although the selected clouds occupy a relatively narrow mass range of $\sim$\,$10^{4.5-5.3}$\,M$_\odot$, they exhibit a remarkable diversity in their physical properties. The cloud radii vary by roughly a factor of three between the smallest and largest cases, and comparable variations are found in v$_{\rm{rel}}$, $\chi$, $\epsilon$ and A$_{\rm{cld}}$. We note that the values of $\chi$ in our sample are lower than those assumed in typical cloud crushing simulations ($\gtrsim$\,$100$; e.g \citealt{scannapieco2015}), although some works have also adopted values as low as $\sim$\,$20$ \citep{schneider2017}. While some clouds are rapidly outflowing, others are infalling toward the halo center, and yet others follow quasi-circular orbits. The galactocentric distances of the clouds span the full extent of the halo, naturally giving rise to a wide range of background density and temperature conditions. Inhomogeneities in the latter two properties are also relatively large, i.e. $\sigma_{\rho, \rm{bck}}$/$\mu_{\rho, \rm{bck}}$ and $\sigma_{\rm{T, bck}}$/$\mu_{\rm{T, bck}}$ are typically of order unity, suggesting that these clouds are embedded in rather non-uniform, multi-phase environments (see also Appendix~\ref{sec:app_inter_iden}).

\begin{figure*}
    \centering
    \includegraphics[width=18cm]{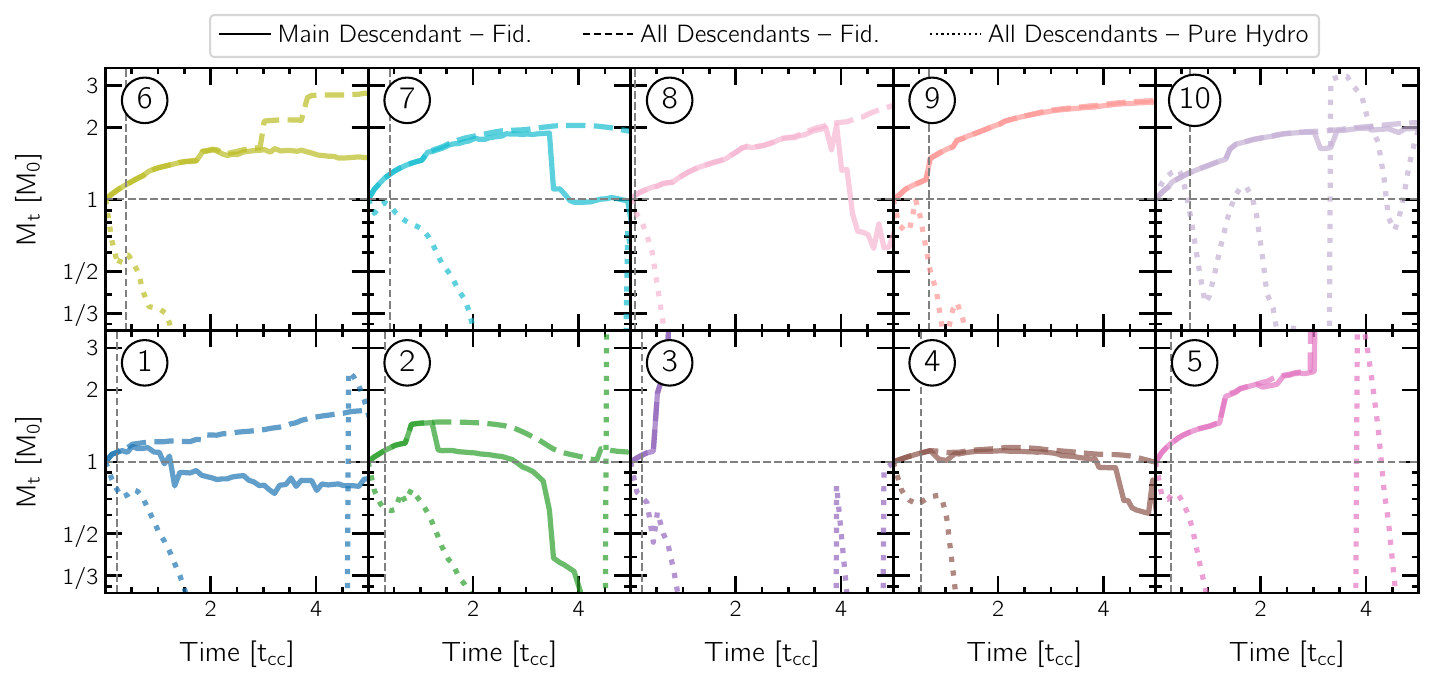}
    \caption{Individual growth profiles for the clouds in our sample. Each panel corresponds to a different cloud, arranged in the same order as Figure~\ref{fig:ic_gallery}. Solid curves show the mass of the main descendant -- the single most massive cloud -- over time, while dashed curves include all descendants, accounting for fragments separated from the original object. Vertical dashed gray lines indicate the median cooling time in the cloud interfaces at the initial conditions. In all cases, this timescale is comparable to or shorter than the cloud crushing timescale (t$_{\rm{cc}}$), indicating that clouds are generally expected to grow steadily in mass. While some clouds follow this expectation, others show more complex behavior: some initially grow before losing mass, while a few undergo major mergers, producing sharp jumps in their masses. For comparison, dotted curves show the mass evolution in a `pure hydrodynamic' simulation, i.e. without the physics of cooling or magnetic fields. In most cases, the mass decays steadily on timescales comparable to t$_{\rm{cc}}$, although abrupt spikes reveal instances where clouds `revive' through mergers with fragments. Interestingly, clouds that merge in the pure hydro case do not always do so in the fiducial case, and vice versa. Overall, the figure illustrates the diversity of cloud growth trajectories and underscores the key role of additional physics in sustaining cloud mass over time.
    }
    \label{fig:mass_v_time}
\end{figure*}

In Figure~\ref{fig:mass_v_time}, we explore the evolution of these clouds over time. Each panel corresponds to a distinct cloud, arranged in the same order as Figure~\ref{fig:ic_gallery}. Solid curves show the mass of the main descendant -- the single most massive cloud -- over time, while dashed curves include all descendants, accounting for fragments separated from the original object. Vertical dashed gray lines indicate the median cooling time, in units of t$_{\rm{cc}}$, in the cloud interfaces at t\,$=$\,$0$ (t$_{\rm{cool,init}}$). For comparison, dotted curves show the mass evolution in a `pure hydrodynamic' simulation, i.e. without the inclusion of physics of cooling or magnetic fields.

In nearly all panels, the dotted curves decline rapidly with time, indicating that clouds are quickly destroyed through interactions with their background media, as extensively suggested for purely adiabatic scenarios \citep[e.g.][]{klein1994}. While some clouds are permanently disrupted, in other cases they re-form through coagulation with neighboring cold gas fragments, as evidenced by the abrupt spikes in the corresponding panels \citep[see also][]{waters2019,gronke2023}.

As expected, the growth profiles turn more complex in the fiducial case, where additional physics is included. In all panels, the mass of the main descendant grows steadily over the first $\sim$\,$\rm{t_{cc}}$, but the subsequent evolution differs markedly across the sample. For instance, clouds $9$ and $10$ undergo sustained, continuous entrainment throughout the full simulation time of $5$\,$\rm{t_{cc}}$, whereas cloud $6$ grows for the first $\sim$\,$3$\,$\rm{t_{cc}}$ before its mass plateaus.
In contrast, several other clouds gradually lose mass over time, ending the simulation with a lower mass at $5$\,$\rm{t_{cc}}$ than at t\,$=$\,$0$. Two clouds ($3$ and $5$) experience mergers that produce abrupt mass increases. Interestingly, clouds that merge in the pure-hydrodynamic case do not always merge in the fiducial case, and vice versa. This arises since the properties of the background medium also change when cooling and magnetic fields are switched off, thereby altering the interactions between the cloud and its surroundings -- the discussion of which we return to later in this section.

For all clouds, t$_{\rm{cool,init}}$ is comparable to or shorter than t$_{\rm{cc}}$, as shown by the vertical dashed gray lines. A number of earlier results with idealized setups suggest that clouds in this regime are expected to grow in mass \cite[e.g.][]{gronke2020,kanjilal2021}. Note, however, that some other works argue that the cooling time of the hot-phase primarily dictates cloud evolution \citep{li2020}. Others suggest that the shear time-scale, R$_{\rm{cld}}$/v$_{\rm{rel}}$, is physically more important than $\rm{t_{cc}}$, as it sets the time over which gas in the interface can be entrained onto the cloud before being swept past \citep{abruzzo2023}.

Although the mass of the main descendant does not generally increase monotonically, the total mass of all descendants (dashed curves) does exhibit steady growth in most cases, even when the main descendant loses mass over time (e.g. cloud $1$). This growth, however, typically approaches a regime of saturation within $5$\,$\rm{t_{cc}}$, in contrast to idealized cloud-crushing setups where the buildup of cold mass continues at roughly the same rate for much longer time-scales \citep[e.g.][]{scannapieco2015,armillotta2016}. Non-homogeneous background wind properties in such simulations may reduce the growth rate, although the total mass can still steadily increase over long timescales \citep{dutta2025}. Finally, there are cases (clouds $2$ and $4$) in which even the total mass of all descendants at $5$\,$\rm{t_{cc}}$ remains roughly the same as at t\,$=$\,$0$, despite t$_{\rm{cool,init}}$\,$\lesssim$\,$\rm{t_{cc}}$, suggesting that additional factors may govern the evolution of these clouds.

\begin{figure*}
    \centering
    \includegraphics[width=18cm]{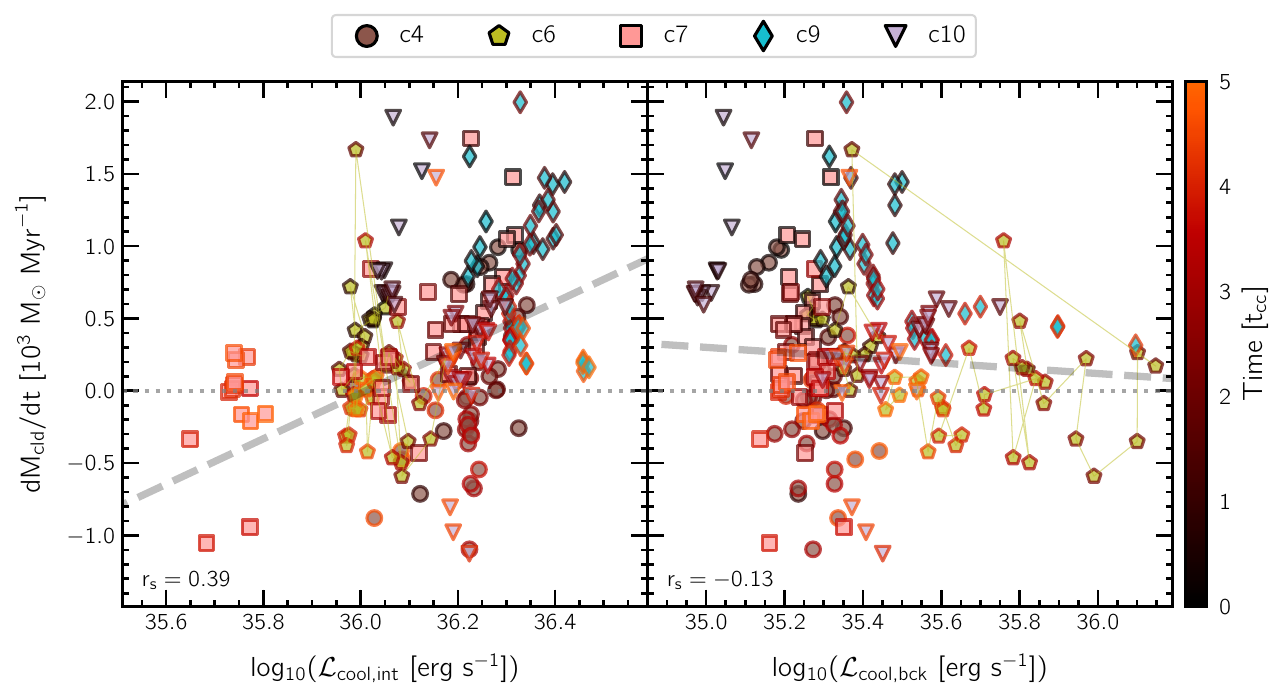}
    \caption{
    Examining the connection between the differential mass growth of clouds and the cooling luminosity of their ambient gas. To avoid overcrowding the panels, we randomly select five clouds, each shown with a unique combination of marker shape and color. The y-axis gives the mass-growth rate of the main descendant between consecutive snapshots, with snapshot time encoded by the marker edge color, as indicated by the colorbar. In the left (right) panel, the x-axis shows the cooling luminosity of gas in the interface (background) layer surrounding the clouds, $\mathcal{L}_{\rm{cool,int}}$ ($\mathcal{L}_{\rm{cool,bck}}$), evaluated at the snapshot corresponding to each point. The horizontal gray dotted line at zero separates net mass growth from net mass loss, the gray dashed curves show the best-fit lines for the stacked cloud sample, and the Spearman correlation co-efficient, r$_{\rm{s}}$, for the stacked point set is shown on the bottom left. At the population level, the differential mass growth increases towards larger values of $\mathcal{L}_{\rm{cool,int}}$, while the trend flattens out in the background. 
    } 
    \label{fig:dm_dt_vs_coolLum}
\end{figure*}

To better interpret the diversity in cloud mass evolution, in Figure~\ref{fig:dm_dt_vs_coolLum}, we investigate the relation between the instantaneous mass growth rate of the main descendants of clouds (y-axis) and the cooling luminosity of the surrounding gas. The left panel shows the cooling luminosity of the interface layer ($\mathcal{L}_{\rm{cool,int}}$), while the right panel shows that of the background  ($\mathcal{L}_{\rm{cool,bck}}$). To avoid overcrowding the panel, we randomly select five clouds, each shown with a unique combination of marker shape and color. The horizontal gray dotted line at zero separates net mass growth from net mass loss. To guide the eye toward the overall directionality of the trend, the gray dashed curves show the best-fit lines for the stacked cloud sample, and the Spearman correlation coefficient (r$_{\rm{s}}$) for the stacked set of points is shown in the bottom left of each panel. To emphasize the non-monotonic evolution of the interface and the background over time for individual clouds, we encode the snapshot time by the marker edge color, as indicated by the colorbar. For one cloud, we further connect the scatter points between neighboring snapshots with a thin curve.

In the left panel, we find that, amidst the non-negligible scatter, the mass growth of clouds correlates with $\mathcal{L}_{\rm{cool,int}}$, with a moderately strong scaling (r$_{\rm{s}}$\,$=$\,$0.39$). Transitioning into the background (right panel), the trend begins to flatten out, revealing a rather constant, although slightly decreasing, mass growth towards higher values of $\mathcal{L}_{\rm{cool,bck}}$, with r$_{\rm{s}}$\,$=$\,$-0.13$. Note that these scalings are largely prevalent only at the cloud population level, and do not necessarily hold for individual clouds at all snapshots.

These results are broadly consistent with a number of past studies which have demonstrated that the long-term survival and evolution of clouds may be sustained, to a large extent, by cooling in the (warm) interface layer \citep[e.g.][]{gronke2018,fielding2020}. Furthermore, our findings suggest that cooling of the hot background gas does not directly aid cold mass growth, contrary to a subset of previous works \citep[e.g.][]{li2020}. Properties of the background, however, may affect cloud growth, as we discuss next.

\begin{figure*}
    \centering
    \includegraphics[width=18cm]{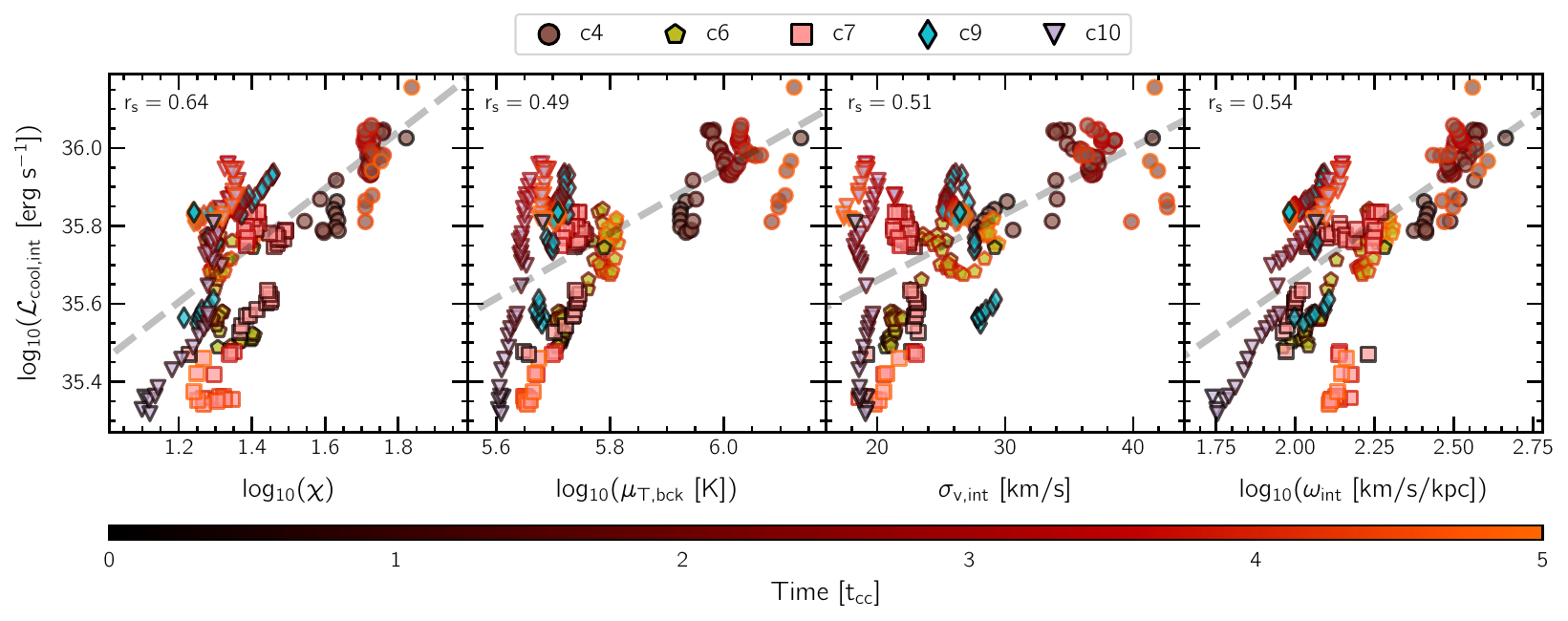}
    \caption{
    Assessing the connection between various physical properties and the cooling luminosity of the interface layer around clouds. From left to right, the x-axes correspond to the cloud overdensity ($\chi$), the mean temperature of the background ($\mu_{\rm{T}, \rm{bck}}$), turbulent velocity ($\sigma_{\rm{v}, \rm{int}}$) and vorticity ($\omega_{\rm{int}}$) in the interface. The y-axis shows the cooling luminosity of the interface ($\mathcal{L}_{\rm{cool,int}}$), while the scatter points and the rest of the panel layout follow the same convention as Figure~\ref{fig:dm_dt_vs_coolLum}. At the cloud population level, broad correlations between $\mathcal{L}_{\rm{cool,int}}$ and all four properties are present, as discussed in more detail in the main text. 
    }
    \label{fig:coolLum_vs_prop}
\end{figure*}

In Figure~\ref{fig:coolLum_vs_prop}, we examine the connection between $\mathcal{L}_{\rm{cool,int}}$ (y-axis) and various physical properties: the cloud overdensity ($\chi$), the mean temperature of the background ($\mu_{\rm{T}, \rm{bck}}$), turbulent velocity ($\sigma_{\rm{v}, \rm{int}}$) and vorticity ($\omega_{\rm{int}}$) in the interface, shown on the x-axis from left to right, respectively. The scatter points and the gray dashed curves follow the same convention as Figure~\ref{fig:dm_dt_vs_coolLum}, and the Spearman correlation coefficients are shown on the top-left of the corresponding panels. 

Overall, at the cloud population level, we find that $\mathcal{L}_{\rm{cool,int}}$ typically increases towards larger values of all four properties shown, with correlations being moderately strong (r$_{\rm{s}}$\,$\sim$\,$0.5-0.6$). Past results suggest that higher values of these physical properties generally correspond to regimes of higher cold gas mass growth.

For instance, \cite{fielding2020} show that the typical inflow velocity of gas into the cloud is expected to increase towards higher values of $\chi$, which effectively yields a higher mass in-flux into the cold phase \citep{tan2024,ramesh2025}. Turbulence may also boost the mass growth of clouds in certain physical regimes, as a result of stretching and increase in surface area \citep{ghosh2025}. 

Although rapid mixing between the cold cloud and the hot background, corresponding to large values of vorticity, may lead to an out-flux of material away from the cloud (\citealt{ramesh2025}; see also \citealt{fielding2022,mohapatra2023}), it simultaneously populates the intermediate temperature range, which may aid in cloud survival via radiative cooling \citep{scannapieco2015,gronke2018}.
Furthermore, hotter backgrounds could shift the peak temperatures of the interface ($T_{\rm{int}} \sim \sqrt{T_{\rm{cold}} T_{\rm{hot}}}$) to regimes where cooling is rapid and efficient \citep{begelman1990}.

In addition to the properties discussed above, we mention that $\mathcal{L}_{\rm{cool,int}}$ also increases towards higher levels of background inhomogeneities ($\sigma_{\rho, \rm{bck}}$/$\mu_{\rho, \rm{bck}}$), albeit with a weaker monotonic trend (r$_{\rm{s}}$\,$=$\,$0.27$). This is broadly consistent with a number of analytic works that predict strong density perturbations may promote condensation of hot gas via run-away cooling (\citealt{field1965}; see also \citealt{sharma2010}, \citealt{dutta2022}), thereby increasing the mass budget of the cold phase. 

Taken together, these scalings indicate that cold-phase growth depends on multiple physical properties that not only differ across clouds but also evolve in distinct ways over time. We thus suggest that the heterogeneous behavior seen in Figure~\ref{fig:mass_v_time} is a result of the complex and individualized initial conditions derived for each system. We caution, however, that each panel of Figure~\ref{fig:coolLum_vs_prop} isolates a single parameter, whereas the evolution likely arises from the combined influence of several coupled processes. The reported correlations therefore should not be interpreted as simple one-to-one relationships. Moreover, although broad trends emerge at the population level, they do not necessarily characterize individual objects at every snapshot.

Our extraction framework thus presents a trade-off. While it moves towards capturing the environmental complexity of a CGM-like medium more realistically than idealized models, this realism complicates causal interpretation: diverse mass-growth histories cannot be readily attributed to a single controlling variable, as is likely also true in reality.

\begin{figure*}
    \centering
    \includegraphics[width=18cm]{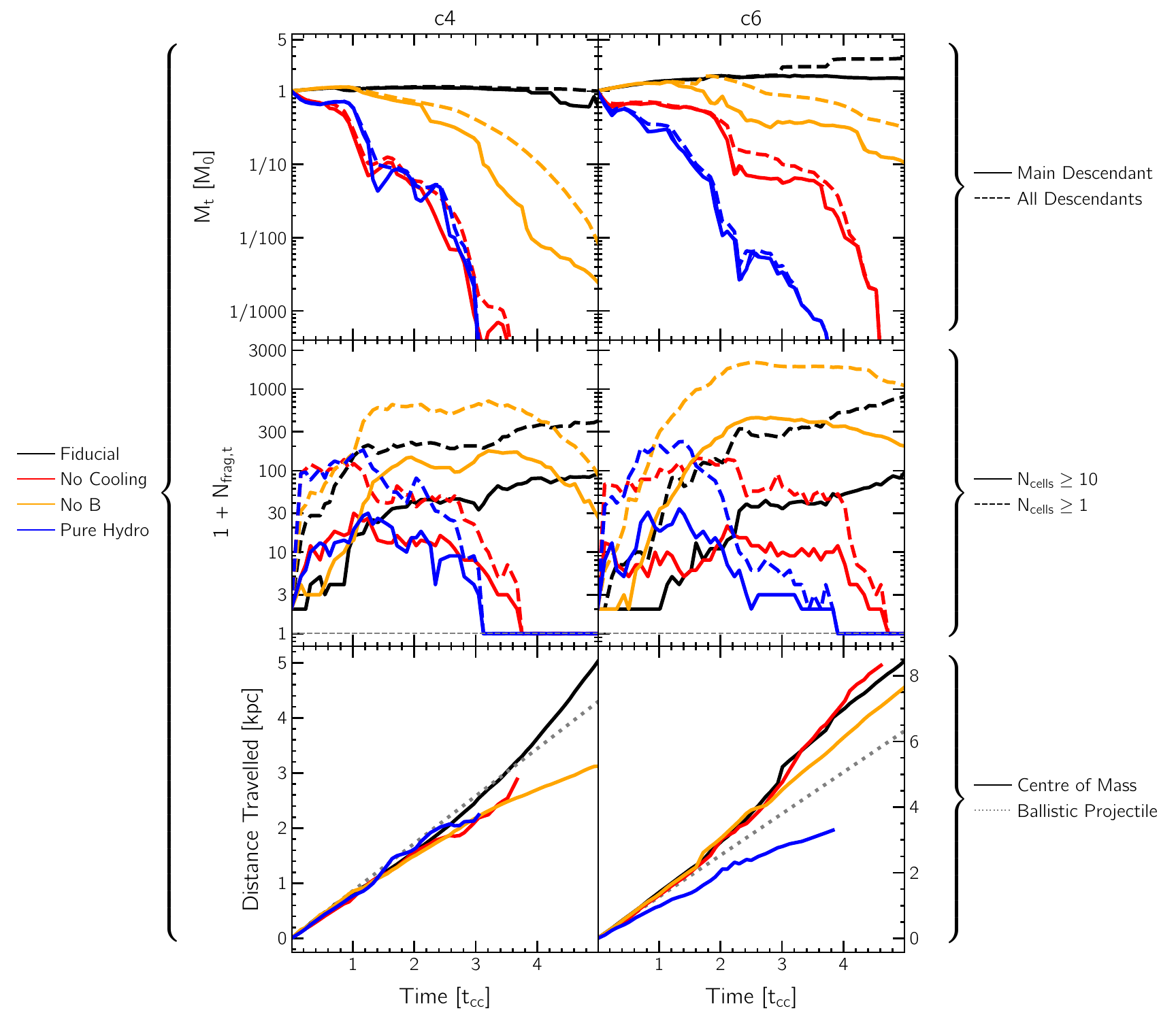}
    \caption{Probing the impact of different physical processes on the evolution of clouds. We focus on two representative examples: cloud \#4 (left column) and cloud \#6 (right; see Table~\ref{table:cloud_props} for details). In all panels, we contrast four physics variation runs: the fiducial setup (black), no radiative cooling (red), no magnetic fields (orange), and a pure hydrodynamical case with both cooling and magnetic fields absent (blue). The upper row shows the mass evolution of clouds over time, with the main descendants plotted as solid curves and all descendants as dotted curves, following the convention in Figure~\ref{fig:mass_v_time}. While both clouds survive in the fiducial case, they lose mass rapidly in the other runs -- especially when cooling is disabled. The middle row shows the number of fragments into which each cloud breaks apart, distinguishing between fragments with at least 10 cells (solid) and those down to the resolution limit of one cell (dashed). Both clouds fragment quickly when either magnetic fields or cooling are disabled, with the absence of magnetic fields leading to a greater level of breakup. The lower row shows the distance traveled by each cloud over the course of the simulation. For a fair comparison, we measure the center-of-mass motion of all fragments (solid curves). The dotted gray line marks the expected distance if the clouds moved ballistically at their initial velocities. Clouds in the fiducial runs travel farther than these ballistic estimates -- an apparently counterintuitive result that reflects the complex background velocity field, where cloud motion is shaped not only by deceleration due to drag but also by acceleration driven by momentum exchange with the ambient medium. Taken together, these results demonstrate that radiative cooling and magnetic fields significantly influence the survival, structure, and motion of CGM clouds.}
    \label{fig:phy_varn_runs}
\end{figure*}

As a final exploration in this work, Figure~\ref{fig:phy_varn_runs} examines the impact of different physical processes on cloud evolution. We focus on two representative examples: clouds \#4 (left column) and \#6 (right; see Table~\ref{table:cloud_props} for details). As detailed in Section~\ref{ssec:phys_proc}, we contrast four physics variation runs: the fiducial setup (black), no radiative cooling (red), no magnetic fields (orange), and the pure hydrodynamical case (blue). Such variations are difficult to interpret in standard cosmological simulations, since excising cooling or magnetic fields ultimately affects galaxy growth and supermassive black hole evolution, which in turn modifies the CGM through altered feedback outputs (see \citealt{pillepich2018} for a detailed discussion). Our approach provides a more controlled framework in this regard, making it possible to probe the direct impact of these physical processes on cloud growth.

In the upper row, we examine the growth of cloud mass over time, contrasting the mass of the main descendant (solid curves) with that of all descendants (dashed curves). For both clouds, turning off cooling leads to an almost immediate drop in mass at t\,$=$\,$0$, seen in both the solid and dashed curves. In contrast, removing magnetic fields produces mass evolution largely similar to the fiducial case over the first $\sim$\,$\rm{t_{cc}}$, with a sharp decline at later times; the total cold mass at $5$\,$\rm{t_{cc}}$ however remains higher than in the no-cooling case. 

Interestingly, the differences between the red and blue curves are minimal for cloud 4 (left) but more pronounced for cloud 6 (right). While not explicitly shown, we note that for cloud $4$ in the no-cooling run, the ratio of magnetic to thermal pressure ($\beta^{-1}$\,$=$\,$\rm{P_B/P_{th}}$) within the main descendant remains roughly constant at $\sim$\,$0.7$ over time, whereas for cloud $6$ it increases from $\sim$\,$0.4$ to $1.2$ over $5$\,$\rm{t_{cc}}$. We suspect that this enhanced magnetic pressure support contributes to the behavior described above \citep[see also][]{nelson2020,fielding2023}. Overall, this highlights the complex impact of magnetic fields on cloud growth, particularly in configurations where the initial field structure is non-trivial.

In the middle row, we plot the the number of fragments (N$_{\rm{frag}}$) into which each cloud breaks apart, distinguishing between cloudlets with at least 10 cells (solid) and those down to the resolution limit of one cell (dashed). In all cases, the solid and dashed curves are qualitatively similar -- i.e. they follow the same overall trends -- but with the dashed curves boosted by a factor of $\sim$\,a few. We therefore focus our discussion on the solid curves hereafter.

In the fiducial case, fragmentation is minimal during the first $\sim$\,$\rm{t_{cc}}$; this is followed by a phase in which N$_{\rm{frag}}$ increases monotonically out to $5$\,$\rm{t_{cc}}$, for both clouds. Similar to the top panel, the run with no magnetic fields (orange) shows early behavior comparable to the fiducial case, after which N$_{\rm{frag}}$ rises rapidly. Unlike the black curves, however, the number of fragments reaches a global maximum within $5$\,$\rm{t_{cc}}$, followed by a decline in N$_{\rm{frag}}$. This occurs due to a combination of (a) the destruction of existing fragments and (b) the main descendant producing fewer new fragments as its own mass rapidly declines, both of which signify the net decline of the cold phase, as also seen in the top panel.

In the two cases where cooling is absent (red and blue), fragmentation proceeds rapidly at early times compared to the fiducial run (black). However, without cooling, neither the main progenitor nor the smaller fragments entrain mass \citep{gronke2018}, leading to a subsequent rapid decline in N$_{\rm{frag}}$, consistent with the abrupt loss of mass seen in the top row, signaling the destruction of the cloud and eventual mixing into the background. Similar to the top panel, the red and blue curves are largely similar for cloud $4$ (left), but diverge in the right panel; as above, we attribute this difference to the enhanced pressure support.

In the bottom row, we show the distance traveled by each cloud over the course of the simulation. Because fragmentation rates differ markedly across the variation runs, we track the center-of-mass motion of all fragments (solid curves) for a fair comparison. This provides a measure of the distance traveled by the surviving portion of the initial cloud at each snapshot. The dotted gray line marks the expected distance if the clouds moved ballistically at their initial velocities.

Interestingly, in the fiducial case for both clouds, the distance traveled initially follows the gray curves but eventually exceeds them. Although this may seem counterintuitive, since clouds are expected to decelerate over time due to drag and therefore move slower than the ballistic prediction\footnote{For cases in which a cloud initially at rest is placed in a background wind with the same relative velocity, it is instead accelerated by ram pressure. However, Galilean invariance dictates that the overall distance traveled remains unchanged.}, this behavior simply reflects the complex velocity field at the start of the simulation, where random, uneven motions in the background induce additional acceleration through momentum exchange, allowing the clouds to travel farther than simple kinematic expectations would suggest. In addition, clouds may be embedded in strong bulk outflows and thus accelerated by these faster background winds \citep{schneider2018,nelson2019}, even if they are not directly interacting with them at the start of the simulation (see also Figure~\ref{fig:propVis}).

In the absence of magnetic fields, the overall distance traveled by both clouds is smaller than in the fiducial run. We attribute this to enhanced fragmentation in the former (middle row), which increases the effective surface area ($\rm{A_{eff}}$) exposed to the background, and consequently the drag force ($\rm{F_d}$\,$\propto$\,$\rm{A_{eff}}$). While the red curve diverges from the black at $\sim$\,$2$\,$\rm{t_{cc}}$ for cloud $4$ (left), the two interestingly remain largely similar in the right panel. As before, this underscores the non-trivial role of the initial orientations of both the magnetic and velocity fields.

We note that the above results of Figure~\ref{fig:phy_varn_runs} exhibit a number of \textit{qualitative} similarities with previous studies using idealized setups. In regimes where radiative cooling is weak, i.e., t$_{\rm{cool,init}}/\rm{t_{cc}}$\,$\gg$\,$1$, including cases where cooling is absent (t$_{\rm{cool,init}}/\rm{t_{cc}}$\,$\sim$\,$\infty$), such simulations also predict a net loss of cold-phase mass over time. However, the rate of mass loss is typically slower\footnote{Note that the two studies referenced here do not include magnetic fields and are thus most comparable to our pure hydro case.}: for example, \cite{marinacci2010} find that in the no-cooling case, their cloud loses only $\sim$\,$2$\,$\%$ of its mass over a timescale of $\rm{t_{cc}}$. In another region of parameter space, \cite{nakamura2006} show that the mass at $5$\,$\rm{t_{cc}}$ ranges from $\sim$\,$0.4$\,$-$\,$0.8$ of the initial mass, depending on the initial configuration. 

By contrast, in our pure hydrodynamical runs, the cold-phase mass is depleted more rapidly, dropping to $\sim$\,$10$\,$\%$ of the initial value by $\sim$\,$\rm{t_{cc}}$ and being effectively destroyed within $\sim$\,$3$\,$-$\,$4$\,$\rm{t_{cc}}$. While $\rm{t_{cc}}$ is not perfectly well defined in our case, as these clouds are neither spherical nor fully characterized by a single $\chi$ or $\rm{v_{rel}}$, and thus the comparison may not truly be apples-to-apples, this alone is unlikely to entirely account for the quantitative difference noted above, especially since clouds in idealized simulations typically decline to $\sim$\,$10$\,$\%$ of their initial mass over much longer timescales of $\gtrsim$\,$10$\,$\rm{t_{cc}}$ \citep{gronke2018}. As with our earlier results, this behavior of rapid mass loss cannot be attributed to any single physical parameter, but instead reflect the interplay of several factors.

Our findings on fragmentation rates over time are also broadly qualitatively consistent with previous studies. For instance, \cite{scannapieco2015} show using simulations with radiative cooling that clouds break into smaller cloudlets along their tails over $\gtrsim$\,$O(\rm{t_{cc}})$, with the number of fragments sensitive to Mach number. Using cloud crushing setups that include ideal MHD, \cite{mccourt2015} find that clouds fragment less when magnetic fields are present, whereas in the absence of fields, the cloud effectively dissolves into a mist of smaller cloudlets over similar time-scales.

The biggest difference between our results and those in the existing literature concerns the distance traveled over time (bottom row). In particular, \cite{mccourt2015} show that a cloud is expected to travel shorter distances in the presence of magnetic fields than in the non-MHD case, at least when field lines drape around the interface, as the associated magnetic tension effectively increases the drag force. We, however, find that this may not always be the case.

We suspect that this is primarily due to the interface magnetic fields not being perfectly draped for the cases considered: at t\,$=$\,$0$, the magnetic curvature at the interfaces is $\kappa_{\rm{int}}$\,$\sim$\,$0.46$ and $0.3$ kpc$^{-1}$ for clouds $4$ and $6$, respectively. For a perfectly draped configuration, one expects $\kappa_{\rm{int}}$\,$\rm{R_{cld}}$\,$\sim$\,$1$ \citep{dursi2008}; for the clouds examined here, $\kappa_{\rm{int}}$\,$\rm{R_{cld}}$ is only $\sim$\,$0.18$ and $0.13$. Moreover, as noted earlier, the motion of clouds in our setup is influenced not only by drag forces but also by accelerations arising from other components. This possibly further contributes to the contrast in results highlighted above, particularly since disabling magnetic fields also alters the properties of the background, producing different effective accelerations (see also Figure~\ref{fig:mass_v_time}). 

\section{Summary and Conclusions}\label{sec:summary}
In this paper, we conduct cloud-scale ($\sim$\,kpc) simulations with initial conditions sampled from a high-resolution cosmological magneto-hydrodynamical zoom-in of a Milky Way-like galaxy. In particular, we select ten distinct circumgalactic clouds of mass M$_{\rm{cld}}$\,$\sim$\,$10^{4.5-5}$\,M$_\odot$ from the $z$\,$=$\,$0$ volume of Project GIBLE \citep{ramesh2024b}, which achieves a mass resolution of $m_{\rm{gas}}$\,$\sim$\,$200$\,M$_\odot$ in the multi-phase CGM. We further boost the local resolution around clouds using a targeted refinement scheme, reaching $m_{\rm{gas}}$\,$\sim$\,$0.2$\,M$_\odot$. At this level, our selected clouds are resolved by $\sim$\,$50-60$ cells per radius, comparable to current high-resolution idealized cloud-crushing simulations. Overall, our approach offers a balance between these simplified setups that achieve high resolution but lack the complexity of the CGM, and standard cosmological runs that more readily produce a self-consistent CGM but at substantially lower resolution.

We find that the mass growth profiles of these ten clouds are all distinct (Figure~\ref{fig:mass_v_time}). Although all reside in the fast cooling regime -- where the cooling time of interface gas (t$_{\rm{cool}}$) is shorter than the cloud-crushing time scale (t$_{\rm{cc}}$) -- only a subset show sustained mass growth. Furthermore, for these clouds, the net mass influx typically saturates within $5$\,t$_{\rm{cc}}$, in contrast to simpler idealized setups that experience prolonged periods of mass growth \citep[e.g.][]{armillotta2016,gronke2018,kanjilal2021}.

The overall evolution of the clouds is intrinsically coupled to their properties, as well as those of the ambient medium. In particular, we find their mass growth to correlate with $\mathcal{L}_{\rm{cool,int}}$, the cooling luminosity of the interface (Figure~\ref{fig:dm_dt_vs_coolLum}). This, in turn, scales with various properties, such as the cloud overdensity ($\chi$), the mean temperature of the background ($\mu_{\rm{T}, \rm{bck}}$), turbulent velocity ($\sigma_{\rm{v}, \rm{int}}$) and vorticity ($\omega_{\rm{int}}$) in the interface (Figure~\ref{fig:coolLum_vs_prop}). We caution, however, that multiple properties of the clouds and the background vary simultaneously, making it challenging to disentangle the influence of the various processes; consequently, these scalings may reflect either direct causal effects or indirect outcomes of this multidimensional interplay.

The inclusion or removal of different physical processes also impacts growth (Figure~\ref{fig:phy_varn_runs}). In our cloud sample, radiative cooling is by far the dominant factor enabling survival, while magnetic fields play a smaller role, particularly by suppressing fragmentation into less massive cloudlets. Interestingly, clouds can travel farther than ballistic projectiles initialized with the same velocities, due to acceleration from momentum exchange with the background velocity field. Overall, this underscores the complex impact of magnetic- and velocity-fields on cloud growth, amongst other physical properties, particularly when the initial configurations are far from idealized.

This work represents the first attempt at simulating cloud-related phenomena evolved in the full $\Lambda$CDM cosmological context at resolutions comparable to modern idealized wind-tunnel setups. Future efforts will expand the cloud sample to achieve a statistically robust ensemble, as well as examine the impact of additional physical processes. For example, previous studies with idealized simulations suggest that thermal conduction can suppress the growth of cold gas \citep{armillotta2017}, or even lead to a net mass loss \citep{marcolini2005,vieser2007}, although certain magnetic field configurations may mitigate these effects \citep[e.g.,][]{bruggen2023}. Whether these findings hold under more realistic, non-trivial initial conditions -- or deviate significantly -- remains an open question. In parallel, the added information on $\sim$\,kpc scales could be incorporated into sub-grid schemes for lower-resolution cosmological simulations, either on-the-fly \citep[e.g.][]{weinberger2023,butsky2024,smith2024} or in post-processing \citep[e.g.][]{hummels2023,nelson2025}, which we reserve for the future.

\section*{Data Availability}
Collaboration on new projects with the GIBLE simulation data, including the extracted cloud volumes, is welcome, and potential collaborators are encouraged to contact the authors directly. Data directly related to this publication is available upon reasonable request to the corresponding author. This work has benefitted from the \texttt{scida} library\footnote{\url{github.com/cbyrohl/scida}} \citep{byrohl2024} for data analysis, and the \texttt{temet} package\footnote{\url{github.com/dnelson/temet}} \citep{nelson2025} for the various visualization figures presented.

\begin{acknowledgements}
RR acknowledges support by the World Premier International Research Center Initiative (WPI), MEXT, Japan. RR and DN acknowledge funding from the Deutsche Forschungsgemeinschaft (DFG, German Research Foundation) through an Emmy Noether Research Group (grant number NE 2441/1-1). This work is supported by the DFG under Germany's Excellence Strategy EXC 2181/1 - 390900948 (the Heidelberg STRUCTURES Excellence Cluster). MB acknowledges funding by the Deutsche Forschungsgemeinschaft (DFG, German Research Foundation) under Germany's Excellence Strategy -- EXC 2121 ``Quantum Universe'' --  390833306. RR thanks Annalisa Pillepich, Prateek Sharma, Enrico Garaldi and Kentaro Nagamine for valuable discussions. This work has made use of the VERA supercomputer of the Max Planck Institute for Astronomy (MPIA) operated by the Max Planck Computational Data Facility (MPCDF), and of NASA's Astrophysics Data System Bibliographic Services.
\end{acknowledgements}

\bibliographystyle{mnras}
\bibliography{references}

\appendix

\section{Resolution Convergence}\label{sec:app_res_conv}
As detailed in Section~\ref{ssec:ref_scheme}, results presented in the main text are derived from simulations where the local resolution around clouds is boosted to $m_{\rm{gas}}$\,$\sim$\,$0.2$\,M$_\odot$. In this appendix, we assess numerical convergence by comparing the evolution of cloud $4$ at three increasingly coarser levels of resolution.

\begin{figure*}
    \centering
    \includegraphics[width=18cm]{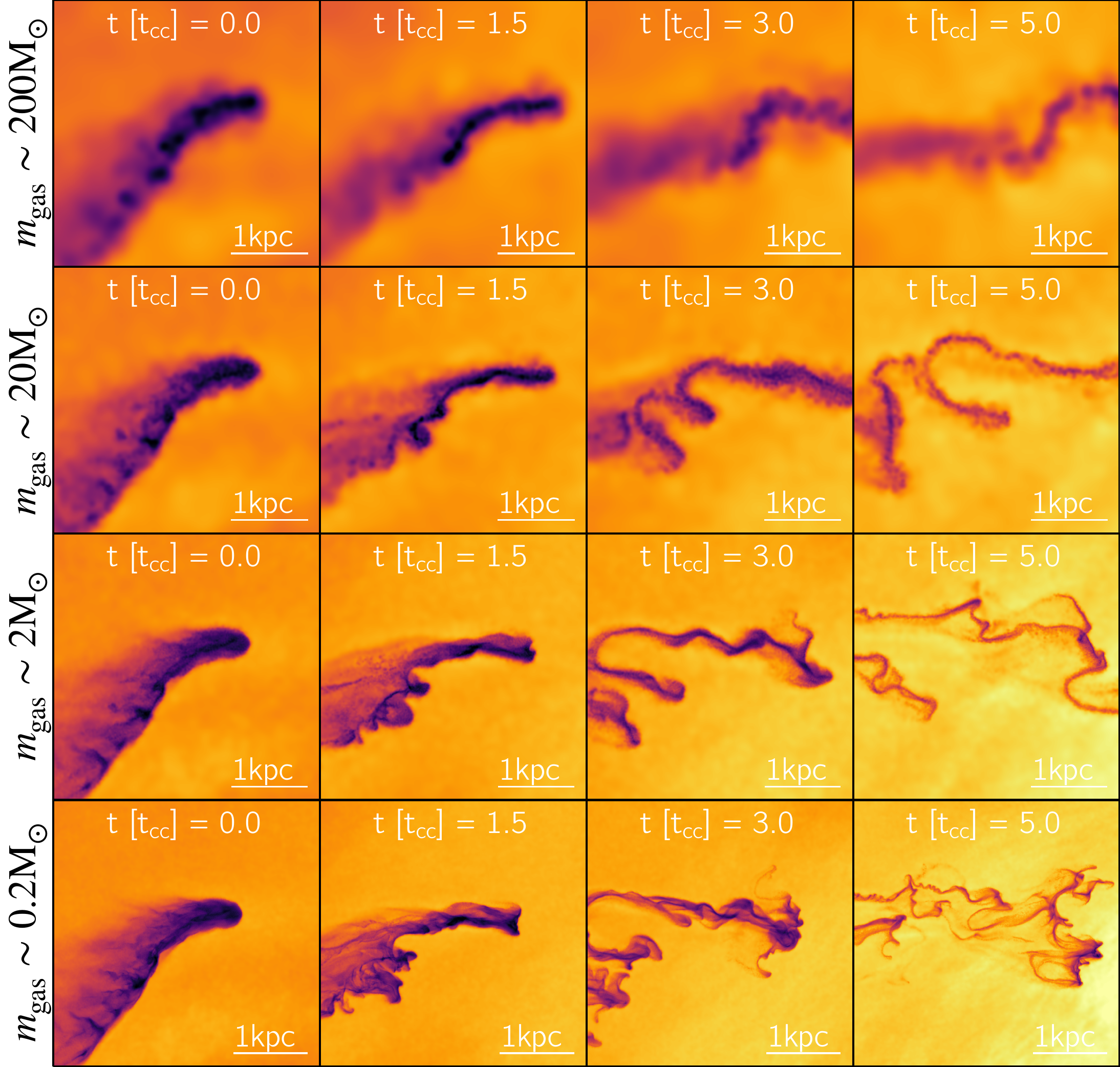}
    \includegraphics[width=7cm]{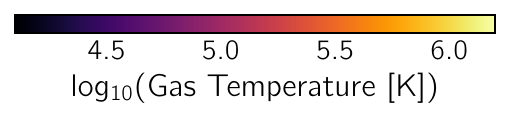}
    \caption{A visual assessment of the impact of numerical resolution on cloud growth (cloud \#4). From top to bottom, panels show runs with $m_{\rm gas}$\,$\sim$\,200\,M$_\odot$, $\sim$\,20\,M$_\odot$, $\sim$\,2\,M$_\odot$ and $\sim$\,0.2\,M$_\odot$, respectively. From left to right, images correspond to times t\,$=$\,$0$, 1.5\,$\rm{t_{cc}}$, 3\,$\rm{t_{cc}}$ and 5\,$\rm{t_{cc}}$, respectively. Colors indicate the mean temperature in a thin $\pm$\,R$_{\rm{cl}}$ projection centered on the main descendant of the cloud. Higher resolution reveals smaller-scale features, as expected, while the overall large-scale morphology of the low-resolution clouds resembles a blurred version of their high-resolution counterparts.}
    \label{fig:app_res_conv_vis}
\end{figure*}

In Figure~\ref{fig:app_res_conv_vis}, we begin with a visual assessment of the impact of numerical resolution on the growth of the selected cloud. From top to bottom, panels show runs with $m_{\rm gas}$\,$\sim$\,200\,M$_\odot$, $\sim$\,20\,M$_\odot$, $\sim$\,2\,M$_\odot$ and $\sim$\,0.2\,M$_\odot$, respectively. From left to right, images correspond to times t\,$=$\,$0$, 1.5\,$\rm{t_{cc}}$, 3\,$\rm{t_{cc}}$ and 5\,$\rm{t_{cc}}$, respectively. Colors indicate the mean temperature in a thin $\pm$\,R$_{\rm{cl}}$ projection centered on the main descendant of the cloud, with the panels oriented such that the mean velocity vector is oriented along the positive x-axis. The higher-resolution runs reveal features on smaller scales, as expected; however, the overall large-scale morphology of the low-resolution clouds resembles a blurred version of their high-resolution counterparts.

\begin{figure}
    \centering
    \includegraphics[width=9cm]{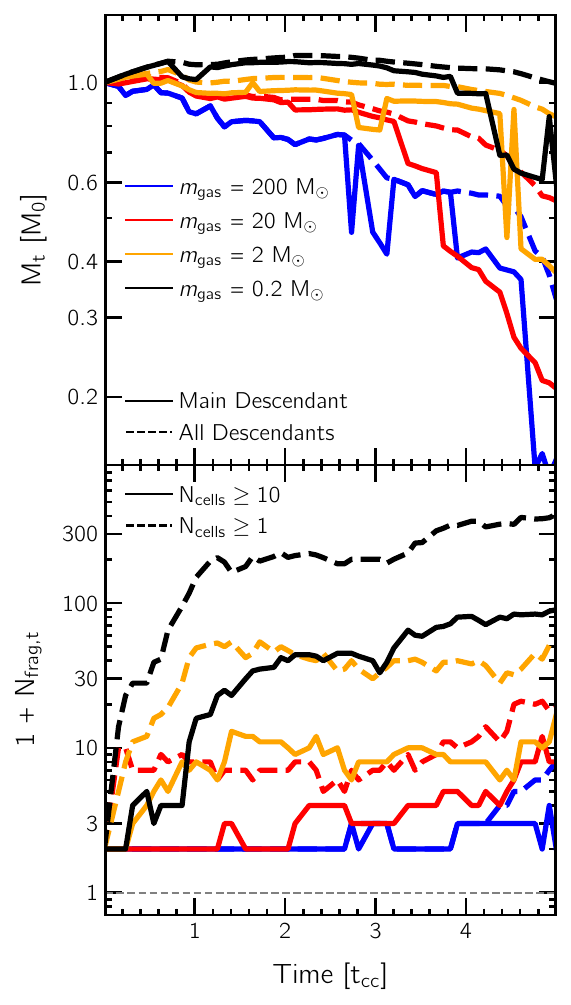}
    \caption{A quantitative assessment of the impact of numerical resolution on cloud mass evolution (upper panel) and fragmentation over time (lower panel) for the same cloud shown in Figure~\ref{fig:app_res_conv_vis}. The blue, red, orange, and black curves correspond to runs with $m_{\rm gas}$\,$\sim$\,200\,M$_\odot$, $\sim$\,20\,M$_\odot$, $\sim$\,2\,M$_\odot$ and $\sim$\,0.2\,M$_\odot$, respectively. In the upper panel, solid curves show the growth of the main descendant, while dashed curves indicate the total mass summed over all descendants. Both trends suggest that clouds retain more mass at higher resolution, with profiles approaching convergence towards $\sim$\,0.2\,M$_\odot$. In the lower panel, solid curves show the number of fragments with N$_{\rm{cells}}$\,$\geq$\,$10$, and dashed curves include all fragments. Increased fragmentation is evident in the high-resolution runs, whereas it is strongly suppressed at low resolution.}
    \label{fig:app_res_conv_plot}
\end{figure}

We turn to a more quantitative exploration in Figure~\ref{fig:app_res_conv_plot}. The blue, red, orange, and black curves correspond to runs with $m_{\rm gas}$\,$\sim$\,200\,M$_\odot$, $\sim$\,20\,M$_\odot$, $\sim$\,2\,M$_\odot$ and $\sim$\,0.2\,M$_\odot$, respectively.

In the top panel, we focus on the mass evolution over time, distinguishing between the main descendant (solid curves) and all descendants (dashed). In both sets of curves, clouds retain more mass at higher resolution \citep[see also][]{goldsmith2016}. Interestingly, and importantly, the growth profiles appear to tend towards convergence at $m_{\rm gas}$\,$\sim$\,2\,M$_\odot$, with the orange and black curves only offset by a factor of a few, as opposed to the blue curves (lowest resolution) that portray markedly different behaviors.

In the bottom panel, we show the number of fragments (N$_{\rm{frag}}$) formed over time, with those composed of N$_{\rm{cells}}$\,$\geq$\,$10$ ($1$) in solid (dashed) curves. Higher resolution gives rise to a larger number of less massive cloudlets, with no sign of convergence \citep{mccourt2018}. Intriguingly, N$_{\rm{frag}}$ increases monotonically for the highest resolution run, while it plateaus for $m_{\rm gas}$\,$\sim$\,2\,M$_\odot$, suggesting that the smallest cloudlets are more likely to survive at higher resolution, ultimately giving rise to the difference between the black and orange curves in the top panel.

\section{Defining Cloud Interfaces and Backgrounds}\label{sec:app_inter_iden}

As mentioned in Section~\ref{ssec:def}, we consider the interface to be the \textit{transition} region between the cloud and the asymptotic background. We here delve deeper into the numerical details of the algorithm.

For a given cloud, we first identify concentric Voronoi \textit{layers} defined through the geometric connectivity of the tessellation. The zeroth layer consists of the cloud cells themselves. Each successive layer is constructed from non-cold cells (T\,$>$\,$10^{4.5}$\,K) that neighbor the preceding shell, excluding any already assigned to earlier layers. This procedure yields a topologically unique result for a given point set, with layers corresponding to increasing graph distance and naturally adapting to the arbitrary shape of the cloud.

\begin{figure*}
    \centering
    \includegraphics[width=18cm]{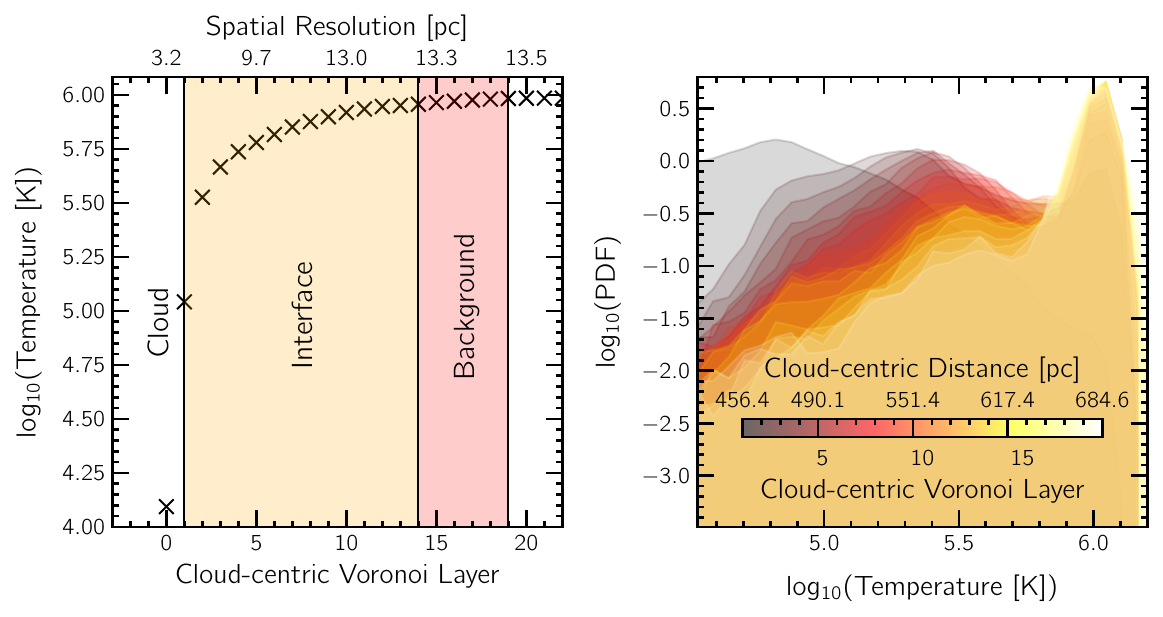}
    \caption{A sample demonstration of our algorithm employed to define cloud interfaces and backgrounds. Following the identification of concentric shells of Voronoi \textit{layers} around a given cloud, we compute the mean temperature of gas cells in successive layers, as shown in the left panel. The interface is thereafter considered to be the \textit{transition} region that extends from the outer shell of the cloud, out to the layer whose temperature is $95\%$ of the asymptotic value. The background is next defined as the region stacked over the five successive layers of gas cells, as shown through the different shaded regions. Note that although we characterize layers through a single mean temperature, they are inherently multi-phase, as shown on the right.}
    \label{fig:app_inter_iden}
\end{figure*}

For each layer, we then compute the mean temperature of all member gas cells. The left panel of Figure~\ref{fig:app_inter_iden} shows this as a function of the index of the Voronoi layer (with the upper x-axis additionally showing the mean radii of gas cells within each layer, as a proxy for the local spatial resolution; see also the right panel for the mean cloud-centric distance of each layer): the mean temperature increases rather monotonically with increasing distance, before asymptoting to a typical hot-phase value (see also \citealt{nelson2020}).

We thereafter define the interface to be the region that extends from the outer shell of the cloud out to the layer whose temperature is $95\%$ of the asymptotic value. The background is next defined as the region stacked over the five successive layers of gas cells. The shaded regions in the left panel of Figure~\ref{fig:app_inter_iden} highlight the result of the above algorithm. 

While this does involve arbitrary choices, i.e. the value of $95\%$ for truncating the transition region and five layers for the background, varying these parameters slightly does not qualitatively change our results in any strong manner. We find this to be the case since (a) the properties of the interface are dictated by the inner layers, given larger number of gas cells due to higher densities (and thus reducing the threshold of $95\%$ to, e.g., $70\%$ only excises a few of the outer layers, whose contributions are sub-dominant), and (b) the properties of the asymptotic region are statistically similar regardless of the number of layers considered. 

Note that although we characterize layers through a single mean temperature, they are inherently multi-phase. We show this in the right panel through a series of temperature PDFs for the various layers, colored by their corresponding index (see also the discussion of Table~\ref{table:cloud_props}). 

\end{document}